\documentclass[12pt,letter]{article}

\usepackage{graphicx, epsfig, color}
\def\to{\rightarrow}

\def\bi{\begin{itemize}}
\def\ei{\end{itemize}}
\def\tG{\tilde G}
\def\te{\tilde e}
\def\ta{\tilde a}

\def\tu{\tilde u}

\def\tf{\tilde f}
\def\td{\tilde d}

\def\tst{\tilde t}

\def\tg{\tilde g}

\def\tell{\tilde\ell}

\def\tw{\widetilde W}
\def\tz{\widetilde Z}
\def\alt{\stackrel{<}{\sim}}
\def\agt{\stackrel{>}{\sim}}
\def\be{\begin{equation}}  
\def\ee{\end{equation}}  
\def\bea{\begin{eqnarray}}  
\def\eea{\end{eqnarray}}  
\def\beas{\begin{eqnarray*}}  
\def\eeas{\end{eqnarray*}}  
\newcommand\prd[3]{{\it Phys.\ Rev.\ }{\bf D #1} (#2) #3}
\newcommand\prep[3]{{\it Phys.\ Rept.\ }{\bf #1} (#2) #3}
\newcommand\prl[3]{{\it Phys.\ Rev.\ Lett.\ }{\bf #1} (#2) #3}
\newcommand\plb[3]{{\it Phys.\ Lett.\ }{\bf B #1} (#2) #3}
\newcommand\jhep[3]{{\it J. High Energy Phys.\ }{\bf #1} (#2) #3}

\newcommand\apj[3]{{\it Astrophys.\ J. }{\bf #1} (#2) #3}
\newcommand\ijmpd[3]{{\it Int.\ J.\ Mod.\ Phys.\ }{\bf D #1} (#2) #3}
\newcommand\npb[3]{{\it Nucl.\ Phys.\ }{\bf B #1} (#2) #3}
\newcommand\epjc[3]{{\it Eur.\ Phys.\ J. }{\bf C #1} (#2) #3}
\newcommand\ptp[3]{{\it Prog.\ Theor.\ Phys.\ }{\bf #1} (#2) #3}
\newcommand\arnps[3]{{\it Ann.\ Rev.\ Nucl.\ Part.\ Sci.}{\bf  #1} (#2) #3}

\newcommand\zpc[3]{{\it Z.\ Phys.\ {\bf C}}{\bf  #1} (#2) #3}

\newcommand\npps[3]{{\it Nucl.\ Phys.\ }{\bf #1} {\it(Proc.\ Suppl.)} (#2) #3}
\newcommand\njp[3]{{\it New\ Jou.\ Phys.}{\bf  #1} (#2) #3}
\newcommand{\hepph}[1]{hep-ph/#1}

\newcommand{\astroph}[1]{astro-ph/#1}

\begin{document}
\begin{titlepage}

\vspace{0.5cm}
\begin{center}
{\Large \bf 
Neutralino, axion and axino cold dark matter\\
in minimal, hypercharged  and gaugino AMSB
}\\ 
\vspace{1.2cm} \renewcommand{\thefootnote}{\fnsymbol{footnote}}
{\large Howard Baer$^{1}$\footnote[1]{Email: baer@nhn.ou.edu },
Radovan Derm\' \i\v sek$^2$\footnote[2]{Email: dermisek@indiana.edu},
Shibi Rajagopalan$^1$\footnote[3]{Email: shibi@nhn.ou.edu},
Heaya Summy$^1$\footnote[4]{Email: heaya@nhn.ou.edu}} \\
\vspace{1.2cm} \renewcommand{\thefootnote}{\arabic{footnote}}
{\it 
1. Dept. of Physics and Astronomy,
University of Oklahoma, Norman, OK 73019, USA \\
2. Dept. of Physics,
Indiana University, Bloomington IN 47405, USA \\
}

\end{center}

\vspace{0.5cm}
\begin{abstract}
\noindent 
Supersymmetric models based on anomaly-mediated SUSY breaking (AMSB)
generally give rise to a neutral wino as a WIMP 
cold dark matter (CDM) candidate,
whose thermal abundance is well below measured values.
Here, we investigate four scenarios to reconcile AMSB dark matter
with the measured abundance:
1. non-thermal wino production due to decays of scalar fields
({\it e.g.} moduli),
2. non-thermal wino production due to decays of gravitinos,
3. non-thermal wino production due to heavy axino decays,
and
4. the case of an axino LSP,
where the bulk of CDM is made up
of axions and thermally produced axinos.
In cases 1 and 2, 
we expect wino CDM to constitute the entire measured DM abundance, 
and we investigate wino-like WIMP direct and indirect detection rates.
Wino direct detection rates can be large, and more importantly, are bounded
from below, so that ton-scale noble liquid detectors should access 
all of parameter space for $m_{\tz_1}\alt 500$ GeV.
Indirect wino detection rates via neutrino telescopes and space-based cosmic
ray detectors can also be large.
In case 3, the DM would consist of an axion plus wino admixture, whose exact 
proportions are very model dependent. In this case, it is possible that both
an axion and a wino-like WIMP could be detected experimentally.
In case 4., we calculate the re-heat temperature of the universe
after inflation.
In this case, no direct or indirect WIMP signals should be seen, 
although direct detection of relic axions may be possible.
For each DM scenario, we show results for the minimal AMSB model, as
well as for the hypercharged and gaugino AMSB models.
\vspace*{0.8cm}


\end{abstract}


\end{titlepage}

\section{Introduction}
\label{sec:intro}

Supersymmetric (SUSY) models of particle physics are very attractive in that they 
stabilize the gauge hierarchy problem, and provide an avenue for the incorporation of
gravity via local SUSY, or supergravity\cite{wss}. 
They also receive some indirect experimental support via the unification of gauge couplings under
Minimal Supersymmetric Standard Model (MSSM) renormalization group evolution (RGE)\cite{drw}, and they
provide several different candidates (neutralinos, gravitinos, axions/axinos, $\cdots$) 
which can serve as cold dark matter (CDM) in the universe.
If evidence for SUSY is found at LHC, then a
paramount question will be: what is the mechanism of SUSY breaking, and how is it communicated
to the visible sector? Some of the possibilities proposed in the 
literature include: gravity-mediation (SUGRA) with a gravitino mass $m_{3/2}\sim 1$ TeV\cite{sugra},
gauge-mediation (GMSB) with $m_{3/2}\ll 1$ TeV\cite{gmsb}, and anomaly mediation (AMSB) with
$m_{3/2}\sim 100$ TeV\cite{rs,glmr,recent}. 
 
Anomaly-mediated supersymmetry breaking (AMSB) models have received much
attention in the literature due to their attractive properties\cite{rs,glmr}:
1. the soft supersymmetry (SUSY) breaking terms are completely 
calculable in terms of just one free parameter (the gravitino mass, $m_{3/2}$),
2. the soft terms are real and flavor invariant, thus solving the SUSY flavor and $CP$ 
problems and 3. the soft terms are actually renormalization group invariant\cite{jj}, 
and can be calculated at any convenient scale choice. 
In order to realize the AMSB set-up,
it was proposed that the hidden sector be ``sequestered'' on a separate brane 
from the observable sector in an extra-dimensional
universe, so that tree-level supergravity breaking terms do not dominate the 
soft term contributions. 
Such a set-up can be realized in brane-worlds, where SUSY breaking takes
place on one brane, with the visible sector residing on a separate brane. 

A further attractive feature of AMSB models arises due to the scale of their gravitino mass.
SUGRA-type models with $m_{3/2}\sim 1$ TeV suffer from the cosmological gravitino problem. There are two
parts to the gravitino problem\cite{gprob}.
1. If the re-heat temperature after inflation $T_R\agt 10^{10}$ GeV, then the high rate of thermal
gravitino production leads to an overabundance of neutralino dark matter\cite{moroi}.
2. Even for lower values of $T_R\sim 10^5-10^{10}$ GeV, thermal production of $\tG$ followed
by late decays to $particle+sparticle$ pairs injects high energy particles into the cosmic
soup during or after BBN, thus disrupting one of the pillars of Big-Bang theory\cite{moroi}.
If $m_{3/2}\agt$ 5 TeV, then the lifetime $\tau_{\tG}$ drops below $0.1-1$ sec,  and gravitino decay
occurs before or at the onset of BBN.
In AMSB models where $m_{3/2}\sim 100$ TeV, the gravitino is much too short-lived to be afflicted by the BBN
bounds. 

In spite of their attractive features, AMSB models suffer from the well-known problem
that slepton mass-squared parameters are found to be negative, giving rise to tachyonic states.
The original ``solution'' to this problem was to posit that scalars acquire as well a 
universal mass $m_0$, which when added to the AMSB SSB terms, renders them positive\cite{rs,glmr}.
The derived form of soft SUSY breaking terms, supplemented by a universal scalar mass $m_0$
and implemented at the GUT scale, constitutes what is usually called the minimal AMSB, 
or mAMSB model. In mAMSB and the additional models described below, it is assumed that
electroweak symmetry is broken radiatively due to the large top quark mass, so that the
magnitude of the $\mu$ parameter is determined to gain the correct value of $M_Z$, and the
bilinear soft term $B$ is traded for the ratio of Higgs field vevs, $\tan\beta$. 

An alternative set-up for AMSB has been advocated in Ref. \cite{dvw},
known as hypercharged anomaly-mediation (HCAMSB).
It is a string-motivated scenario which uses a similar setup as the one 
envisioned for AMSB.
In HCAMSB, the MSSM resides on a D-brane, and the hypercharge gaugino mass
is generated in a geometrically separated hidden sector.
An additional contribution to the $U(1)_Y$ gaugino mass $M_1$ is generated, and 
its magnitude is parametrized by an additional parameter $\alpha$. 
The large value of $M_1$ feeds into slepton mass evolution through the MSSM RGE, and acts 
to lift the weak-scale slepton soft masses beyond tachyonic values.
Thus, the HCAMSB model naturally solves the tachyonic slepton mass problem 
which is endemic to pure AMSB scenarios.

A third scenario has recently been proposed in Ref. \cite{inoamsb}, under the name
gaugino AMSB, or inoAMSB. The inoAMSB model is suggested by recent work on the phenomenology 
of flux compactified type IIB string theory\cite{shanta}, which reduces to $N=1$
supergravity below the compactification scale. 
The essential features of this scenario are that the  gaugino masses are of the 
anomaly-mediated SUSY breaking (AMSB) form, while
scalar and trilinear soft SUSY breaking terms are highly suppressed: they are
taken as $m_0=A_0\simeq 0$ at energy scale $Q\sim M_{GUT}$, at first approximation. 
The normally large value of $M_1$ as generated in AMSB models feeds into the
scalar soft term evolution, lifting slepton soft masses 
to generate an allowable sparticle mass spectrum, 
while at the same time avoiding tachyonic sleptons or charged LSPs
(lightest SUSY particles).
Charged LSPs are common in models with 
negligible soft scalar masses, such as no-scale\cite{noscale} or gaugino mediation models\cite{inoMSB}. 
Since scalar and trilinear soft terms are highly suppressed, the 
SUSY induced flavor and $CP$-violating processes are also suppressed in inoAMSB.

All three of these models-- mAMSB, HCAMSB and inoAMSB-- share the common feature that
the lightest MSSM particle is a neutral wino, while the lightest chargino is wino-like with
a mass $m_{\tw_1}\sim m_{\tz_1}$. The $\tw_1$-$\tz_1$ mass gap is of order $\sim 200$ MeV\cite{matchev}, 
so that dominantly $\tw_1^\pm\to\tz_1\pi^\pm$, with the decay-produced pion(s) being very soft.
The small mass gap makes the $\tw_1$ rather long lived ($\tau_{\tw_1}\sim 10^{-9}$ sec), and it may  
yield observable highly ionizing tracks (HITs) of order $cm$ in length at LHC detectors\cite{amsb_coll}.

An important consequence of wino-like neutralinos is that the thermal abundance of 
neutralino cold dark matter falls generally an order of magnitude or so below
the measured abundance: 
\be
\Omega_{CDM}h^2=0.1123\pm 0.0035\ \ \ 68\%\ CL
\ee
according to the WMAP7 data analysis\cite{wmap7}.
This latter fact has led many to consider AMSB-like models as perhaps less interesting
than SUGRA-type models, wherein the bino-like or mixed bino-higgsino neutralino can more easily 
yield the measured relic abundance.

In this paper, we address the question of the dark matter abundance in AMSB models.
While the calculated thermal abundance of wino-like neutralinos is found to be
below measured values (for $m_{\tz_1}\alt 800$ GeV), 
we find that there exists a variety of attractive methods to augment the wino abundance, 
thus bringing the calculated abundance into accord with experiment.
These include:
\begin{enumerate}
\item Decay of scalar ({\it e.g.} moduli) fields into sparticles, 
ultimately terminating in $\tz_1$ production\cite{mr}.
In this case, the LSP is expected to be a relic wino-like neutralino,
which would constitute {\it all } of the CDM.
\item Thermal production\cite{bbs,ps} of gravitinos $\tG$ and also possibly gravitino production 
via moduli\cite{kyy} or inflaton\cite{ety} decay, 
followed by $\tG\to particle+\ sparticle\to \tz_1 + X$ (where $X=$ assorted SM debris). Here also, the LSP would be a relic wino-like neutralino,
which would constitute all of the CDM.
\item Thermal production of heavy axinos\cite{bs} followed by $\ta\to particle +\ sparticle\to\tz_1 +X$\cite{ckls}. Here, the LSP is again 
a relic wino-like neutralino, but the CDM would consist of a wino-like WIMP
plus axion mixture.
\item A scenario where $m_{\ta}<m_{\tz_1}$, so the $\ta$ 
is instead the stable LSP\cite{ckkr}. In this case, 
a combination of thermally produced axinos plus vacuum mis-alignment produced axions would 
constitute the CDM\cite{bs_ou,bbs_ou}.
\end{enumerate}

In Sec. \ref{sec:models}, we present some details of the three AMSB models which we investigate.
In Sec. \ref{sec:dm}, we present four methods of reconciling the AMSB CDM relic abundance with the
measured value.
Given cases 1 and 2, and possibly 3, where we expect all (or some fraction) of the measured
abundance to consist of relic winos, in Sec. \ref{sec:winodet} we present rates for direct and indirect
detection of wino-like neutralinos. 
Unlike SUGRA models, the wino CDM direct detection
rate is bounded from below. We find the current experiments like Xenon-100
should be able to explore the parameter space of AMSB-like models
with a wino-like neutralino up to $m_{\tz_1}\alt 200$ GeV. 
Next generation detectors such as ton-scale
noble liquids or SuperCDMS should be able to push to $m_{\tz_1}\sim 500$ GeV.
This would correspond to a reach in $m_{\tg}\sim 3850$ GeV, {\it i.e.} well beyond 
the projected reach of LHC.
We also find excellent prospects for indirect detection of wino-like CDM
via detection of wino annihilation
into $\gamma$s, $e^+$s, $\bar{p}$s or $\bar{D}$s in  the galactic halo.
In fact, Kane {\it et al.} have already proposed wino CDM as an explanation for the 
recent anomalies seen by Pamela, ATIC, Fermi and others\cite{kane}. 
Neutrino telescopes such as IceCube will also have a reach for wino-like
neutralinos, especially for large $\tan\beta$.

In case 4, we would expect the axino $\ta$ to be the LSP, and so here no
direct or indirect WIMP detection signals are expected. However, it may be the case that
large amounts of {\it axions} $a$ are produced in the early universe, in which case
direct detection of axions may be possible at ADMX\cite{admx}. 
We present parameter expectations in Sec. \ref{ssec:ata} for the scenario of mixed axion/axino CDM
to occur.
In Sec. \ref{sec:conclude}, we present a summary and our conclusions.

\section{Overview of mAMSB, HCAMSB and inoAMSB models}
\label{sec:models}

\subsection{Minimal AMSB}

The AMSB contribution to the gaugino mass is given by,  
\be
M_i={\beta_{g_i} \over g_i}m_{\frac{3}{2}}= \frac{b_i g_i^2}{16\pi^2}m_{3/2},
\label{eq:amsb1}
\ee
where $\beta_i$ is the corresponding beta function, defined by
$\beta_{g_i}\equiv dg_i/d\ln\mu$. 
The constants $b_i=(33/5,1,-3)$ for $i=1-3$.

The AMSB contribution  to soft SUSY breaking scalar masses 
is given by,
\be
m_{\tf}^2=-{1\over 4}\left\{{{d\gamma}\over{dg}}\beta_g+
{{d\gamma}\over{df}}\beta_f\right\}m_{\frac{3}{2}}^2
\label{eq:amsb2}
\ee
where $\beta_f$ is the $\beta$-function for the corresponding superpotential
Yukawa coupling, and 
$\gamma=\partial\ln Z/\partial\ln\mu$, with $Z$
the wave function renormalization constant.  
Complete expressions for the MSSM can be found {\it e.g.} in Ref's \cite{wss,amsb_coll}.
Since these give rise to tachyonic slepton masses, each term is supplemented by
$+m_0^2$, where $m_0$ is some additional universal contribution to scalar masses.

Finally, the anomaly-mediated
contribution to the trilinear SUSY breaking scalar coupling is given by,
\be
A_f={\beta_f \over f} m_{\frac{3}{2}},
\label{eq:amsb3}
\ee
where $f$ labels the appropriate Yukawa coupling ({\it e.g.} $f_t$ is the top-quark Yukawa coupling).

Thus, the parameter space of the ``minimal'' AMSB model (mAMSB) is given by\cite{amsb_coll}
\be
m_0,\ m_{3/2},\ \tan\beta ,\ sign(\mu ) \ \ \ \ ({\rm mAMSB}).
\ee
In the mAMSB model, we assume as usual that electroweak symmetry is broken radiatively
by the large top-quark Yukawa coupling. Then  the SSB $B$ term and the superpotential $\mu$
term are given as usual by the scalar potential minimization conditions which 
emerge from requiring an appropriate breakdown of electroweak symmetry. 

The above expressions for the soft SUSY breaking terms are usually imposed as GUT-scale
boundary conditions, and weak scale values are calculated via renormalization group evolution.

\subsection{Hypercharged AMSB}

In HCAMSB, SUSY breaking is localized at the bottom of a strongly warped hidden region,
geometrically separated from the visible region where the MSSM resides.
The warping suppresses contributions due to tree-level gravity mediation\cite{Kachru:2007xp} 
and the anomaly mediation can become the dominant 
source of SUSY breaking in the visible sector. 
Possible exceptions to this sequestering mechanism are 
gaugino masses of $U(1)$ gauge symmetries~\cite{rrform}. 
Thus, in the MSSM, the mass of the bino-- the gaugino  of $U(1)_Y$-- can be the 
only soft SUSY breaking parameter not determined by anomaly mediation\cite{dvw}. 
Depending on its size, 
the bino mass $M_1$ can lead to a small perturbation to the spectrum of anomaly mediation, 
or it can be the largest soft SUSY breaking parameter in the visible sector: 
as a result of RG evolution, its effect on other soft SUSY breaking parameters can 
dominate the contribution from anomaly mediation.

We parametrize the HCAMSB SSB contribution $\tilde{M}_1$ using a dimensionless quantity 
$\alpha$ such that $\tilde{M}_1 =\alpha m_{3/2}$; then, $\alpha$ 
governs the size of the hypercharge
contribution to soft terms relative to the AMSB contribution\cite{hcamsb}. 
The soft SUSY breaking terms are then exactly the same as in mAMSB, except there
is no $m_0^2$ contribution to scalar masses, and the $U(1)_Y$ gaugino mass is given by 
\be
M_1 =\left(\alpha+\frac{b_1g_1^2}{16\pi^2}\right) m_{3/2} ,
\ee
so that $\alpha=0$ takes us back to pure AMSB soft terms, with their concommitant
tachyonic sleptons.
Then the parameter space of HCAMSB models is given by
\be
\alpha ,\ m_{3/2},\ \tan\beta ,\ sign(\mu ) \ \ \ \ ({\rm HCAMSB}),
\ee
where the dimensionless $\alpha$ typically ranges between $0.01-0.2$ for allowable spectra\cite{hcamsb}.

\begin{figure}[htbp]
\begin{center}
\includegraphics[angle=-90,width=0.75\textwidth]{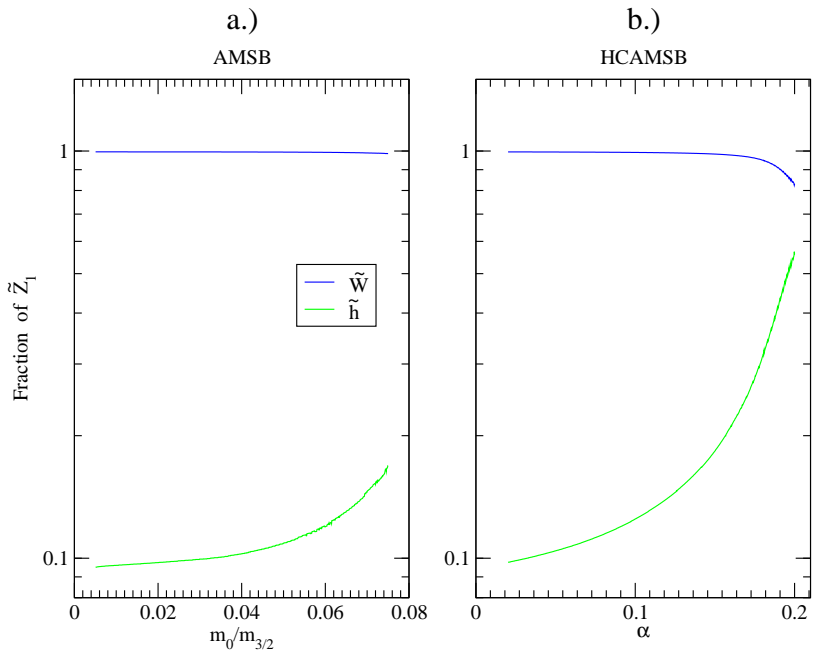}
\vspace{-.5cm}
\caption{The wino and higgsino content of the neutralino $\tz_1$ in {\it a}). mAMSB
versus $m_0$ and {\it b}). HCAMSB versus $\alpha$ for $m_{3/2}=50$ TeV, $\tan\beta =10$ and $\mu >0$.
}
\label{fig:z1frac}
\end{center}
\end{figure}

While the lightest neutralino is mainly wino-like in mAMSB and HCAMSB models, it is important
phenomenologically to realize that for certain regions of parameter space, the $\tz_1$
picks up a substantial higgsino component, thus becoming a mixed wino/higgsino particle. 
This occurs in mAMSB at large $m_0$ values, and in HCAMSB at large $\alpha$ values. 
The situation is shown in Fig. \ref{fig:z1frac} for $m_{3/2}=50$ TeV, $\tan\beta =10$ and $\mu >0$, 
where we show $R_{\tilde W}\equiv v_3^{(1)}$
(the wino component of $\tz_1$ in the notation of Ref. \cite{wss}) and $R_{\tilde h}\equiv\sqrt{v_1^{(1)2}+v_2^{(1)2}}$
(the higgsino component of $\tz_1$) as a function of {\it a}). $m_0$ in mAMSB and {\it b}). $\alpha$ for HCAMSB. 
We see indeed that in both these cases, $R_{\tilde h}$ is increasing as the relevant parameter increases.

\subsection{Gaugino AMSB}

Phenomenologically viable versions of string theory require the stabilization
of all moduli fields as well as weak to intermediate scale supersymmetry
breaking. Models satisfying these criteria were first developed in
the context of type IIB string theory using flux compactifications
and non-perturbative effects on Calabi-Yau orientifolds (CYO's)\cite{Douglas:2006es}. 
The low energy limit of type-IIB string theory after compactification on a CYO
is expected to be $N=1$ supergravity (SUGRA).

Two classes of the above models which yield an interesting supersymmetry breaking scenario have been studied:
\begin{itemize}
\item [{a)}] Those with only a single K\"ahler modulus (SKM models). These are  essentially
of the KKLT type \cite{Kachru:2003aw} but with uplift coming from one-loop quantum effects. 
\item [{b)}] Large Volume Scenario (LVS)\cite{Balasubramanian:2005zx}
models which require at least two moduli. 
\end{itemize}
In both of these types of models, the moduli fields are stabilized
using a combination of fluxes and non-perturbative effects. Additionally,
supersymmetry is broken by the moduli fields acquiring non-zero F-terms
and interacting gravitationally with the MSSM. For both models, the
gauginos acquire mass predominately through the Weyl anomaly
while the classical contribution to the scalar masses and trilinear
coupling constants are naturally suppressed. Here, we take the limit
where scalar and trilinear soft breaking parameters are exactly zero at the
GUT scale: $m_0=A_0=0$, while gaugino masses are of the AMSB form.
The parameter space of inoAMSB models is then given by
\be
m_{3/2},\ \tan\beta ,\ sign(\mu ) \ \ \ \ ({\rm inoAMSB}).
\ee
As shown in Ref. \cite{inoamsb}, 
the inoAMSB model solves the problem of tachyonic scalars in AMSB, since now
the GUT scale scalar masses vanish. It 
also solves the problem of charged LSPs which is endemic to no-scale SUGRA 
or gaugino-mediated SUSY breaking (inoMSB) models, 
which have $m_0=A_0=0$ but with universal gaugino masses equal to $m_{1/2}$.
For inoAMSB models--with scalar masses $m_0=0$ at the GUT scale-- the large
GUT-scale $U(1)_Y$ gaugino mass $M_1$ pulls all scalar
masses to large values, leaving no tachyons and a wino-like neutralino 
as the lightest MSSM particle.

\subsection{Thermally produced wino CDM in AMSB models}

The above mAMSB and HCAMSB models have been included into the Isasugra subprogram of the 
event generator Isajet\cite{isajet}. 
In addition, sparticle mass spectra for the inoAMSB model can easily be generated
using usual mSUGRA input parameters with $m_0=A_0=0$, but with non-universal gaugino masses as
specified by AMSB models.

After input of mAMSB, HCAMSB or inoAMSB parameters, Isasugra
then implements an iterative procedure of solving the MSSM RGEs for the
26 coupled renormalization group equations, taking the weak scale 
measured gauge couplings and third generation Yukawa couplings as inputs, as well
as the above-listed GUT scale SSB terms. Isasugra implements full 2-loop RG running
in the $\overline{DR}$ scheme, and minimizes the RG-improved 1-loop effective
potential at an optimized scale choice $Q=\sqrt{m_{\tst_L}m_{\tst_R}}$\cite{hh} 
to determine the magnitude of $\mu$ and $m_A$. All physical sparticle masses
are computed with complete 1-loop corrections, and 1-loop weak scale threshold corrections
are implemented for the $t$, $b$ and $\tau$ Yukawa couplings\cite{pbmz}. The off-set of the 
weak scale boundary conditions due to threshold corrections (which depend on the entire
superparticle mass spectrum), necessitates an iterative up-down RG running solution.
The resulting superparticle mass spectrum is typically in close accord with other
sparticle spectrum generators\cite{kraml}.

Once the weak scale sparticle mass spectrum is known, then sparticle 
annihilation cross sections may be computed.
To evaluate the thermally produced neutralino relic density, we adopt the
IsaReD program\cite{isared}, which is based on CalcHEP\cite{comphep} to
compute the several thousands of neutralino annihilation and
co-annihilation Feynman diagrams. Relativistic thermal averaging of the
cross section times velocity is performed\cite{gg}.

\begin{figure}[htbp]
\begin{center}
\includegraphics[angle=-90,width=0.75\textwidth]{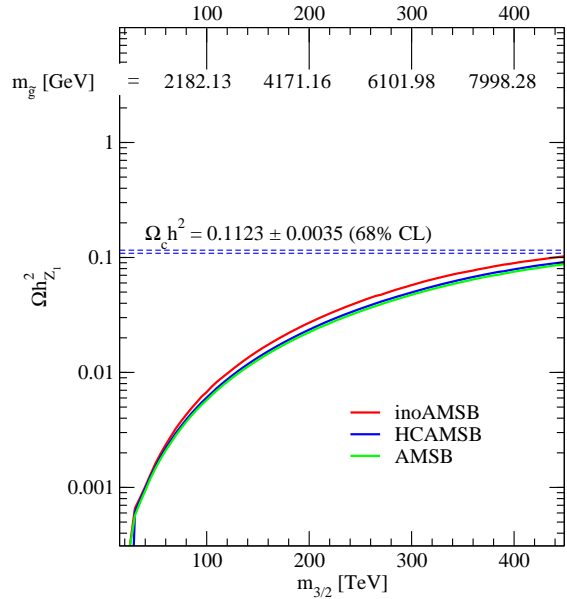}
\vspace{-.5cm}
\caption{Thermally produced relic abundance of wino-like
neutralino cold dark matter in mAMSB, HCAMSB and inoAMSB
versus $m_{3/2}$ for $\tan\beta =10$, with $\mu >0$ and
$m_t=172.6$ GeV. 
For mAMSB, we also take $m_0=0.01 m_{3/2}$ GeV, 
and for HCAMSB, we take $\alpha =0.02$.
}
\label{fig:wino_rd}
\end{center}
\end{figure}

As an example, in Fig. \ref{fig:wino_rd}, we show the thermally produced 
neutralino relic density $\Omega_{\tz_1}h^2$ versus $m_{3/2}$ for all 
three models: mAMSB, HCAMSB and inoAMSB. We take $\tan\beta =10$ and $\mu >0$.
For mAMSB, we also take $m_0=0.01 m_{3/2}$, 
and for HCAMSB, we take $\alpha =0.02$.
On the upper axis, we also indicate the corresponding values of $m_{\tg}$.
We see that the relic abundance is typically well below WMAP7-measured levels, 
until $m_{3/2}\agt 450$ TeV, corresponding to $m_{\tg}\agt 8$ TeV, and
$m_{\tz_1}\agt 1.3$ TeV: well beyond any conceivable LHC reach. 
The well-known tiny relic abundance arises due to the large
$\tz_1\tz_1\to W^+W^-$ annihilation and also $\tz_1\tw_1$ 
and $\tw_1\tw_1$ co-annihilation processes. 
%

\section{Dark matter scenarios for AMSB models}
\label{sec:dm}

\subsection{Neutralino production via moduli decay}

Shortly after the introduction of AMSB models, Moroi and Randall
proposed a solution to the AMSB dark matter problem based on
augmented neutralino production via the decays  of moduli fields
in the early universe\cite{mr}. The idea here is that string theory is 
replete with additional moduli fields: neutral scalar fields with 
gravitational couplings to matter. In generic supergravity theories, the moduli
fields are expected to have masses comparable to $m_{3/2}$.
When the Hubble expansion rate becomes comparable to the moduli
mass $m_\phi$, then an effective potential will turn on, and the moduli
field(s) will oscillate about their minima, producing massive excitations,
which will then decay to all allowed modes: {\it e.g.} gauge boson pairs,
higgs boson pairs, gravitino pairs, $\cdots$. The neutralino 
production rate via moduli decay has been estimated in Ref. \cite{mr}.
It is noted in Ref. \cite{kane2} that the abundance-- given by
\be
\Omega_{\tz_1}h^2\sim 0.1\times\left(\frac{m_{\tz_1}}{100\ {\rm GeV}}\right)
\left(\frac{10.75}{g_*}\right)^{1/4}\left(\frac{\sigma_0}{\langle\sigma v\rangle}\right)
\left(\frac{100\ {\rm TeV}}{m_\phi}\right)^{3/2}
\ee
with $\sigma_0=3\times 10^{-24}$ cm$^3$/sec-- 
yields nearly the measured dark matter abundance for wino-like
neutralino annihilation cross sections and $m_{\phi}\sim 100$ TeV.\footnote{In inoAMSB models, 
we expect moduli with SUSY breaking scale masses, $m_\phi\sim m_{3/2}/\sqrt{V}\ll m_{3/2}$,
where $V$ is the (large) volume of the compactified manifold: $V\sim 10^5$ in Planck units.
In this case, the mechanism would not so easily apply.}
These authors dub this the ``non-thermal WIMP miracle''.

A necessary condition for augmented neutralino production via scalar field decay is that the
re-heat temperature of radiation $T_R$ induced by moduli decays is bounded by $T_R\agt 5$ MeV 
(in order to sustain Big Bang Nucleosynthesis (BBN) as we know it), and $T_R< T_{fo}$, where
$T_{fo}$ is the freeze-out temperature for thermal neutralino production $T_{fo}\sim m_{\tz_1}/20$.
If $T_R$ exceeds $T_{fo}$, then the decay-produced neutralinos will thermalize, and the abundance
will be given by the thermal calculation as usual.

This ``low re-heat'' neutralino production mechanism has been investigated extensively by
Gondolo and Gelmini\cite{gg06}. The low re-heat neutralino abundance calculation depends on the
input value of $T_R$ and the ratio $b/m_\phi$, where $b$ is the average number of neutralinos
produced in moduli decay, and $m_\phi$ is the scalar field mass. 
They note that theories with an underabundance of thermally produced neutralino CDM 
with $\Omega_{\tz_1}^{TP}\agt 10^{-5}\left(\frac{100\ {\rm GeV}}{m_{\tz_1}}\right)$ 
can always be brought into accord with the measured DM abundance for at least one and sometimes
two values of $T_R$.\footnote{Ref. \cite{gg06} also shows that an overabundance of thermally produced
neutralino CDM can also be brought into accord with the measured abundance via dilution of the neutralino
number density by entropy injection from the $\phi$ field decay. Since this case doesn't attain in AMSB
models (unless $m_{\tz_1}\agt 1300$ GeV), we will neglect it here.} 

While the low $T_R\sim 10-1000$ MeV scenario with DM generation via scalar field decay is compelling, 
we note here that it is also consistent with some baryogenesis mechisms: {\it e.g.} Affleck-Dine 
baryogenesis wherein a large baryon asymmetry is generated early on, only to be diluted to observable 
levels via moduli decay\cite{kawa},
or a scenario wherein the baryon asymmetry is actually generated by the moduli decay\cite{kitano}.

\subsection{Neutralino production via gravitino decay}

An alternative possibility for augmenting the production of wino-like neutralinos in AMSB models
is via gravitino production and decay in the early universe. While gravitinos would not be in thermal
equilibrium during or after re-heat, they still can be produced thermally via radiation off ordinary
sparticle scattering reactions in the early universe. The relic density of thermally produced gravitinos
as calculated in Ref's \cite{bbb,ps} is given by
\be
\Omega_{\tG}^{TP}h^2=\sum_{i=1}^{3}\omega_ig_i^2\left(1+\frac{M_i^2}{3m_{3/2}^2}\right)
\log\left(\frac{k_i}{g_i}\right)\left(\frac{m_{3/2}}{100\ {\rm GeV}}\right)\left(\frac{T_R}{10^{10}\ {\rm GeV}}\right),
\ee
where $g_i$ and $M_i$ are the gauge couplings and gaugino masses evaluated at scale $Q=T_R$, and
$\omega_i=(0.018,0.044,0.117)$ and $k_i=(1.266,1.312,1.271)$.
Each gravitino ultimately cascade decays down to the wino-like $\tz_1$ state, so the neutralino
relic density is given by
\be
\Omega_{\tz_1}h^2 =\Omega_{\tz_1}^{TP}h^2 +\frac{m_{\tz_1}}{m_{3/2}}\Omega_{\tG}^{TP}h^2 .
\label{eq:gravo}
\ee

\begin{figure}[htbp]
\begin{center}
\includegraphics[angle=-90,width=0.75\textwidth]{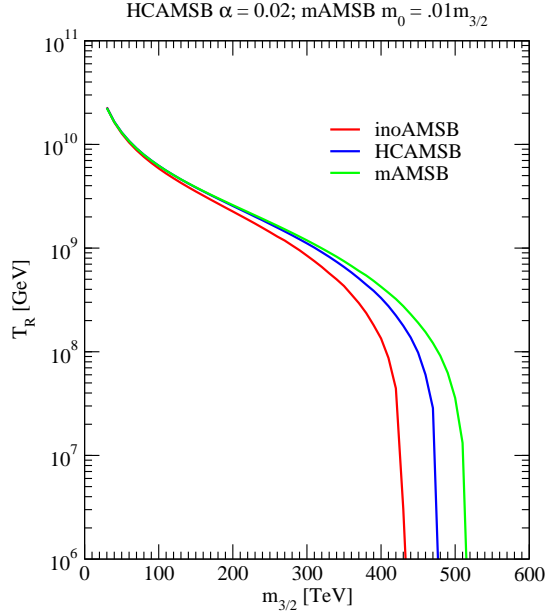}
\vspace{-.5cm}
\caption{
Plot of allowed region of $T_R\ vs.\ m_{3/2}$ plane allowed
for wino-like neutralino DM from thermal production plus 
thermally produced gravitino decay.
}
\label{fig:tr_m32}
\end{center}
\end{figure}

A plot of the value of $T_R$ and $m_{3/2}$ which is required to yield $\Omega_{\tz_1}h^2=0.11$ 
from Eq'n \ref{eq:gravo} is shown in Fig. \ref{fig:tr_m32}
for mAMSB ($m_0=0.01m_{3/2}$), HCAMSB ($\alpha=0.02$) and inoAMSB using $\tan\beta =10$ and $\mu >0$.
The region above the  $\Omega_{\tz_1}h^2=0.11$ curves would yield too much dark matter, 
while the region below the curves yields too little.

We should consider the curves shown in Fig. \ref{fig:tr_m32} as only indicative of the simplest scenario
for wino production via gravitino decay. Three other effects can substantially change the above picture
from what is presented in Eq. \ref{eq:gravo}. 
\begin{itemize}
\item On the one hand, if moduli fields $\phi_m$
exist with mass $m_{\phi_m}>2m_{3/2}$, then gravitinos can also be produced via moduli production
and decay\cite{kyy}. The exact abundance of these moduli-produced gravitinos is very model dependent, 
and depends on the moduli and gravitino mass and branching fractions.
\item A second case arises if we consider gravitino production via inflaton decay at the end of inflation\cite{ety}.
This production mechanism depends on unknown properties of the inflaton: {\it e.g.} its mass and branching fractions,
and the re-heat temperature generated by inflaton decay.
These latter quantities are very model dependent.
\item Additional entropy production generated via the inflaton, moduli and gravitino decays may also dilute the
above relic abundance in Eq. \ref{eq:gravo}.
\end{itemize}
We will bear in mind that these possibilities permit much
lower or much higher values of $T_R$ and $m_{3/2}$ than those shown by the $\Omega_{\tz_1}h^2=0.1$ 
contour of Fig. \ref{fig:tr_m32}.

\subsection{Neutralino production via heavy axino decay}
\label{ssec:haxino}

A third mechanism for increasing the wino-like relic abundance is presented in Ref. \cite{ckls}, in the 
context of the PQMSSM. If we adopt the Peccei-Quinn (PQ) solution to the strong $CP$ problem within the context
of supersymmetric models, then it is appropriate to work with the PQ-augmented MSSM, which contains
in addition to the usual MSSM states, the  axion $a$, the $R$-parity even saxion field $s$, and the
spin-${1\over 2}$ $R$-parity odd axino $\ta$. The axino can serve as the lightest SUSY particle if it is
lighter than the lightest $R$-odd MSSM particle. The $a$ and $\ta$ have couplings to matter which are
suppressed by the value of the PQ breaking scale $f_a$, usually considered to be in the range 
$10^9\ {\rm GeV}\alt f_a\alt 10^{12}$ GeV\cite{axion_review}.

In Ref. \cite{ckls}, it is assumed that $m_{\ta}>m_{\tz_1}$, where $\tz_1$ is the LSP. In the AMSB scenarios
considered here, we will assume $T_R\alt 10^{10}$ GeV, so as to avoid overproduction of dark matter via gravitinos.
With these low values of $T_R$, we are also below the axino decoupling temperature 
$T_{\ta -dcp}=10^{11}\ {\rm GeV}\left(\frac{f_a}{10^{12}\ {\rm GeV}}\right)^2\left(\frac{0.1}{\alpha_s}\right)^3$, 
so the axinos are never considered at thermal equilibrium\cite{axino}. 
However, axinos can still be produced thermally via radiation off
usual MSSM scattering processes at high temperatures. The calculation of the thermally produced axino abundance, 
from the hard thermal loop approximation, yields\cite{bs}
\be
\Omega_{\ta}^{TP}=h^2\simeq 5.5 g_s^6\ln\left(\frac{1.211}{g_s}\right)
\left(\frac{10^{11}\ {\rm GeV}}{f_a/N}\right)^2
\left(\frac{m_{\ta}}{0.1\ {\rm GeV}}\right)
\left(\frac{T_R}{10^4\ {\rm GeV}}\right)
\label{eq:Oh2_TP}
\ee
where $g_s$ is the strong coupling evaluated at $Q=T_R$ and $N$ is the
model dependent color anomaly of the PQ symmetry, of order 1.
Since these axinos are assumed quite heavy, they will decay to $g\tg$ or $\tz_i\gamma$ modes, 
which further decay until the stable LSP state, assumed here to be the neutral wino, is reached.

If the temperature of radiation due to axino decay ($T_D$) exceeds the neutralino
freeze-out temperature $T_{fo}$, then the thermal wino abundance is unaffected by axino decay.
If $T_D<T_{fo}$, then the axino decay will {\it add} to the neutralino abundance. However, 
this situation breaks up into two possibilities: {\it a}). a case wherein the axinos can dominate the energy 
density of the universe, wherein extra entropy production from heavy axino decay 
may dilute the thermal abundance of the wino-like LSPs, and {\it b}). a case where they don't.
In addition, if the yield of winos from axino decay is high enough, then additional 
annihilation of winos after axino decay may occur; this case is handled by explicit solution of
the Boltzmann equation for the wino number density. 
Along with a component of wino-like neutralino CDM, there will of course
be some component of vacuum mis-alignment produced axion CDM: thus, in this scenario, we expect
a WIMP/axion mixture of CDM.

\subsection{Mixed axion/axino CDM in AMSB models}
\label{ssec:ata}

In this case, we again consider the PQMSSM, as in Subsec. \ref{ssec:haxino}. But now,
we consider a light axino with $m_{\ta}<m_{\tz_1}$, so that $\ta$ is the stable 
LSP\cite{ckkr}.
Here, the thermally produced wino-like neutralinos will decay via $\tz_1\to \ta\gamma$, so we
will obtain a very slight dark matter abundance from neutralino decay:
$\Omega_{\ta}^{NTP}=\frac{m_{\ta}}{m_{\tz_1}}\Omega_{\tz_1}h^2$, since each thermally produced neutralino
gives rise to one non-thermally produced (NTP) axino. We will also produce axinos thermally via
Eq'n \ref{eq:Oh2_TP}. Finally, we will also produce axion CDM via the vacuum mis-alignment mechanism\cite{absik}:
$\Omega_a h^2\simeq \frac{1}{4}\left(\frac{f_a/N}{10^{12}\ {\rm GeV}}\right)^{7/6}\theta_i^2$ (we will
take here the initial mis-alignment angle $\theta_i\simeq 1$). The entire CDM abundance is then the sum 
\be
\Omega_{a\ta}h^2=\Omega_{\ta}^{NTP}h^2+\Omega_{\ta}^{TP}h^2+\Omega_{a}h^2 .
\ee
In this case, the TP axinos constitute CDM as long as $m_{\ta}\agt 0.1$ MeV. The NTP axinos
constitute warm DM for $m_{\ta}\alt 1$ GeV\cite{jlm}, but since their abundance is tiny, this
fact is largely irrelevant. The entire CDM abundance then depends on the parameters
$f_a$, $m_{\ta}$ and $T_R$; it also depends extremely weakly on $\Omega_{\tz_1}h^2$, since this is
usually small in AMSB models.

\begin{figure}[htbp]
\begin{center}
\includegraphics[angle=-90,width=0.75\textwidth]{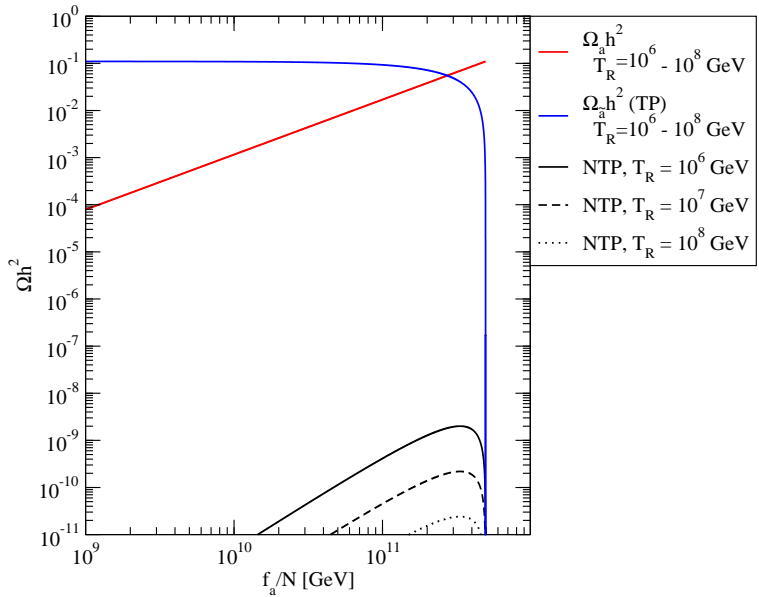}
\vspace{-.5cm}
\caption{Abundance of TP and  NTP axino DM and vacuum-misalignment
production of axion CDM versus $f_a/N$, for various values of $T_R$.
}
\label{fig:oh2_ata}
\end{center}
\end{figure}

As an example, we plot in Fig. \ref{fig:oh2_ata} the three components of mixed axion/axino DM abundance 
from HCAMSB benchmark point 1 in Ref. \cite{hcamsb}:
$\alpha =0.025$, $m_{3/2}=50$ TeV, $\tan\beta =10$ and $\mu >0$. The neutralino thermal DM abundance would be
$\Omega_{\tz_1}h^2=0.0015$ if the $\tz_1$ was stable. 
We require instead $\Omega_{a\ta}h^2 =0.11$, and plot the three components
of $\Omega_{a\ta}h^2$ versus $f_a/N$, for three values of $T_R=10^6$, $10^7$ and $10^8$ GeV. The value of
$m_{\ta}$ is determined by the constraint $\Omega_{a\ta}h^2=0.11$. We see that at low values of 
$f_a/N$, the NTP axino abundance is indeed tiny. Also the axion abundance is tiny since the assumed
initial axion field strength is low. The TP axino abundance dominates. As $f_a/N$ increases, 
the axion abundance increases, taking an ever greater share of the measured DM abundance. The TP axino abundance
drops with increasing $f_a/N$, since the effective axino coupling constant is decreasing. Around
$f_a/N\sim 3\times 10^{11}$ GeV, the axion abundance becomes dominant. It is in this range that 
ADMX\cite{admx} would stand a good chance of measuring an axion signal using their microwave cavity experiment. 

\begin{figure}[htbp]
\begin{center}
\includegraphics[angle=-90,width=0.75\textwidth]{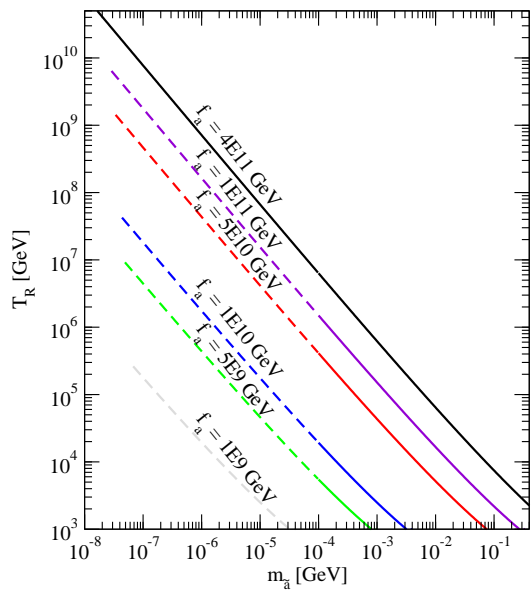}
\vspace{-.5cm}
\caption{Plot of $T_R$ needed to ensure
$\Omega_{a\ta}h^2=0.1$ for HCAMSB benchmark Pt. 1,
versus $m_{\ta}$ for various values of the PQ breaking scale $f_a$.
The dashed curves yield mainly warm axino DM, while solid curves yield
mainly cold mixed axion/axino DM. 
}
\label{fig:trvsmax}
\end{center}
\end{figure}

In Fig. \ref{fig:trvsmax}, we again require $\Omega_{a\ta}h^2=0.11$ for HCAMSB benchmark point 1,
but this time plot the value of $T_R$ which is needed versus $m_{\ta}$, for various 
values of $f_a/N$. The plots terminate at high $T_R$ in order to avoid reaching the axion
decoupling temperature $T_{a-dcp}$. 
Dashed curves indicate regions where over 50\% of the DM is warm, instead of cold. Solid curves
yield the bulk of DM as being cold.

We see that for very light axino masses, and large values of $f_a$, the value of 
$T_R$ easily reaches beyond $10^6$ GeV, while maintaining the bulk of dark matter as cold . 
Such high values of $T_R$ are good enough to sustain
baryogenesis via non-thermal leptogenesis\cite{ntlepto}, although thermal leptogenesis requires
$T_R\agt 10^{10}$ GeV\cite{bdp}. Since $f_a$ is quite large, we would expect that the dominant portion of
DM is composed of relic axions, rather than axinos; as such, detection of the relic axions may be possible at
ADMX\cite{admx}.
While Fig's \ref{fig:oh2_ata} and \ref{fig:trvsmax} were created for the HCAMSB model, 
quite similar results are obtained for the mAMSB or inoAMSB models.
%

\section{Direct and indirect detection of wino CDM in AMSB models}
\label{sec:winodet}

For AMSB dark matter cases 1 and 2 above, it is expected that
the thermal wino abundance will be supplemented by either moduli
or gravitino decay in the early universe, thus increasing the
wino abundance into accord with measured values. 
In these cases, it may be possible to detect relic wino-like WIMPs
with either direct or indirect detection experiments\cite{ullio}.
Also, in case 3 above, it is 
expected that the DM abundance is comprised of an axion/wino mixture. 
If the wino component of this mixture is substantial, then again
direct or indirect WIMP detection may be possible, while if 
axions are dominant, then a WIMP signal is less likely, but direct detection of
relic axions is more likely: in a nearly equal mixture of WIMPs and axions,
possibly detection of both could occur! In case 4 above,
we would expect no WIMP signals to occur in either direct or indirect 
detection experiments.

\subsection{Direct wino detection rates in AMSB models}

Direct detection of WIMPs depends on the WIMP-nucleon
scattering cross section, but also on assumptions about the local 
WIMP density (usually assumed to be $\rho_{local}\simeq 0.3\ {\rm GeV/cm}^3$),
and the velocity distribution of the relic WIMPs (usually assumed to follow
a Maxwellian distribution $f(v)\sim v^2e^{-v^2/v_0^2}$ where $v_0\sim 220$ km/sec, 
the sun's velocity about the galactic center). 
In our case, where WIMPs are
mainly produced non-thermally via moduli, gravitino or axino decay,
the original velocity distribution due to decays will be red-shifted away 
and the current distribution will arise mainly from gravitational infall, as is the
case with thermal WIMP production.
The direct detection reach plots are usually presented in terms of the
WIMP-nucleon scattering cross section. Then, the experimental reach
depends on factors like the mass and spin of the nuclear target,
and the assumed local WIMP density and velocity profiles. 

Direct detection of WIMPs is usually broken down into two components:
detection via spin-independent (SI) interactions ,
and detection via spin-dependent interactions (SD). 
For SI interactions, it may be best to use heavy target nuclei, since
the SI nucleon-WIMP interactions sum coherently over the nuclear mass.
For the SI WIMP-nucleon cross section, we use the Isatools subroutine
IsaReS\cite{bbbo}.

\begin{figure}[htbp]
\begin{center}
\includegraphics[angle=-90,width=0.75\textwidth]{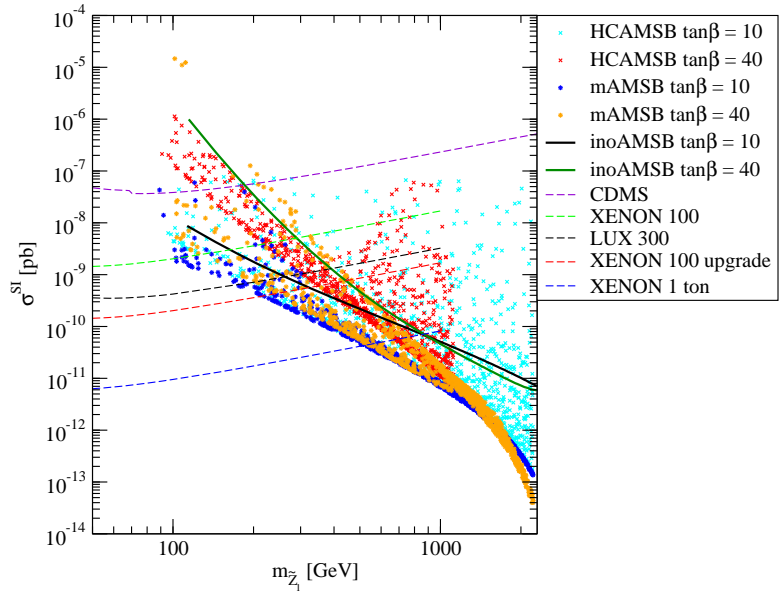}
\vspace{-.5cm}
\caption{Spin-independent $\tz_1-p$ scattering cross section versus
$m_{\tz_1}$ for mAMSB, HCAMSB and inoAMSB models
for $\tan\beta =10$ and 40 and $\mu >0$. The parameters $m_{3/2}$ 
and also $m_0$ (for mAMSB) and  $\alpha$ (for HCAMSB) have been scanned over.
We also show the CDMS limit and projected Xenon and LUX sensitivities.
}
\label{fig:SIDD}
\end{center}
\end{figure}

Our results for SI direct detection of wino-WIMPs is shown in Fig. \ref{fig:SIDD},
in the $\sigma (\tz_1 p)\ vs.\ m_{\tz_1}$ plane.
Here, we scan over $m_{3/2}$ for all models, and $m_0$ (for mAMSB)
and $\alpha$ (for HCAMSB). We show results for $\tan\beta =10$ and 40, while taking $\mu >0$. 
The inoAMSB results occur as lines, since there is no $m_0$ or $\alpha$ dependence.
We keep only solutions that obey the LEP2 limit on a wino-like chargino: $m_{\tw_1}>91.9$ GeV\cite{lepw1lim}.

Several crucial features emerge from the plot.
First, we note that for a given value of $m_{\tz_1}$, the value of
$\sigma (\tz_1 p)$ is bounded from below, unlike the case of the mSUGRA model.
That means that wino-WIMP dark matter can be either detected or excluded for a given
$m_{\tz_1}$ value. 
Second, we note that the cross section values generally fall in the range that is
detectable at present or future DD experiments. The purple contour, for instance, 
exhibits the CDMS reach based on 2004-2009 data, and already excludes some points, 
especially those at large $\tan\beta$. We also show the reach of Xenon-100, LUX, 
Xenon-100 upgrade, and Xenon 1 ton\cite{dm_limits}. 
These experiments should be able to either discover or exclude AMSB models
with $m_{\tz_1}$ values below $\sim 90,\ 100,\ 200$ and 500 GeV respectively.
These WIMP masses correspond to values of $m_{\tg}\sim 690,\ 770,\ 1540$ and 3850 GeV, 
respectively! The latter reach far exceeds the 100 fb$^{-1}$ of 
integrated luminosity reach of LHC for $m_{\tg}$.\footnote{In Ref. \cite{hcamsb}, the 100 fb$^{-1}$
reach of LHC for HCAMSB is found to be $m_{\tg}\sim 2.2-2.4$ TeV. In Ref. \cite{inoamsb}, the
100 fb$^{-1}$ reach of LHC for inoAMSB was found to be $m_{\tg}<2.6$ TeV.} 
For inoAMSB models, where the minimal value of $\sigma^{SI}(\tz_1 p)$ exceeds that of
mAMSB or HCAMSB for a given $m_{\tz_1}$ value, 
the Xenon 1 ton reach is to $m_{\tz_1}\sim 800$ GeV, corresponding to a reach in $m_{\tg}$ of 6200 GeV!

\begin{figure}[htbp]
\begin{center}
\includegraphics[angle=-90,width=0.75\textwidth]{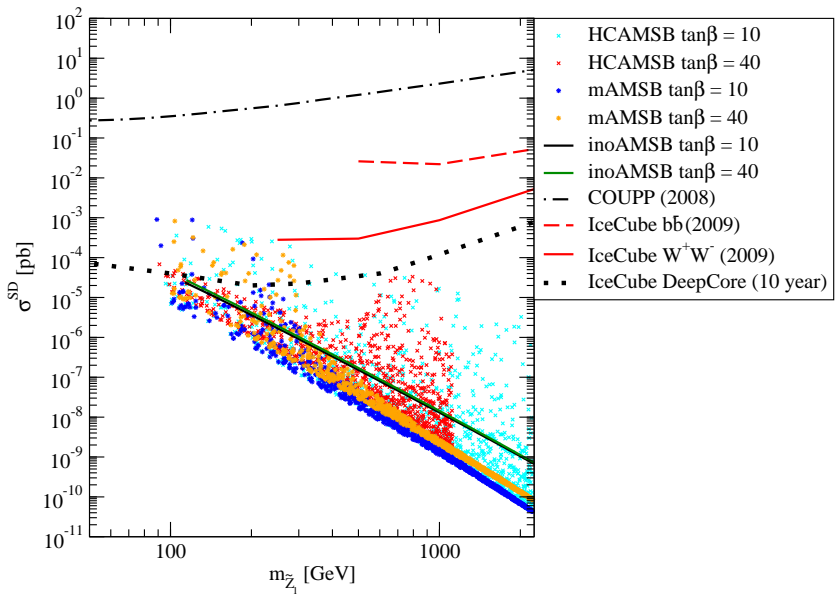}
\vspace{-.5cm}
\caption{Spin-dependent $\tz_1-p$ scattering cross section versus
$m_{\tz_1}$ for mAMSB,  HCAMSB and inoAMSB models
for $\tan\beta =10$ and 40 and $\mu >0$. The parameters $m_{3/2}$ 
and also $m_0$ (for mAMSB) and  $\alpha$ (for HCAMSB) have been scanned over.
We also show the COUPP and IceCube limits in $\sigma^{SD}(\tz_1 p)$.
}
\label{fig:SDDD}
\end{center}
\end{figure}

In Fig. \ref{fig:SDDD}, we show the SD direct detection cross section $\sigma^{SD}(\tz_1 p)$
versus $m_{\tz_1}$ for mAMSB, HCAMSB and inoAMSB models with $\tan\beta =10$ and 40.
We also show a recent limit on this cross section from the COUPP experiment, which is above
the theory expectation by two orders of magnitude. We also show two limits from IceCube in 2009, 
which do approach the theory region, but only for rather large values of $m_{\tz_1}$.
The IceCube SD reach is quite significant, because the rate for WIMP annihilation in the core of the sun
mainly depends on the sun's ability to sweep up neutralinos as it passes along its orbit. The target
here is the solar hydrogen, where the SD cross section usually dominates the SI one, 
since the atomic mass is minimal (an enhancement by number of nucleons per nucleus is usually necessary to make
the SI cross section competetive with the SD one). Since IceCube is mainly sensitive to very high energy muons with
$E_\mu >50$ GeV, it can access mainly higher values of $m_{\tz_1}$. The IceCube DeepCore reach is also shown. The DeepCore
project will allow IceCube to access much lower energy muons, and thus make it more useful for generic WIMP
searches. While DeepCore will access a portion of parameter space, it will not reach the lower limit on
SD cross sections as predicted by AMSB models.

\subsection{Indirect wino detection rates in mAMSB}

Next, we present rates for indirect detection (ID) of wino-like DM via neutrino 
telescopes, and via detection of gamma rays and anti-matter from WIMP annihilation
in the galactic halo.
The ID detection rates depend (quadratically\cite{cdm_reviews}) 
on the assumed galactic DM density (halo) profile. 
We will show results using two profiles: isothermal and Navarro-Frenk-White (NFW)\cite{nfw}
(see {\it e.g.} \cite{njp1} for plots of several recent halo profiles).
Most halo models are in near accord at the earth's
position at $\sim 8$ kpc from the galactic center. However,
predictions for the DM density near the galactic center differ wildly,
which translates to large uncertainties for DM annihilation rates
near the galactic core. The corresponding uncertainty
will be smaller for anti-protons, and smaller still for positrons; since
these particles gradually lose energy while propagating through the
galaxy, they  can reach us only from limited distances over which
the halo density is relatively well-known.
Possible clumping of DM yields an additional source of uncertainty in ID detection rates. 

\begin{figure}[htbp]
\begin{center}
\includegraphics[angle=0,width=0.75\textwidth]{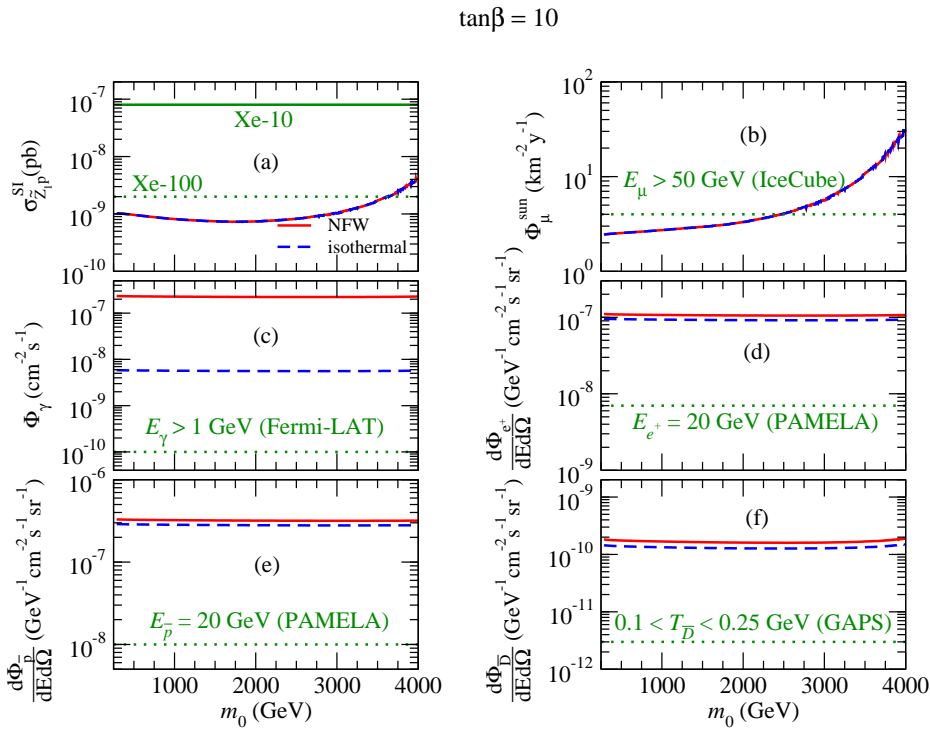}
\vspace{-.5cm}
\caption{Direct and indirect detection rates of neutralino CDM in mAMSB
vs. $m_0$, for $m_{3/2}=50$ TeV, $\tan\beta =10$ and $\mu >0$.
}
\label{fig:idd_m0_tb10}
\end{center}
\end{figure}

In Fig. \ref{fig:idd_m0_tb10}{\it a}.), we show for comparison the
SI direct detection scattering cross section versus $m_0$ in the mAMSB model
for $m_{3/2}=50$ TeV and $\tan\beta =10$. For these parameters, the
wino-like neutralino has mass $m_{\tz_1}\simeq 144$ GeV. We also indicate an 
approximate reach of Xenon-10 and Xenon-100. While the SI direct detection cross section is 
just below Xenon-100 reach for low $m_0$, as $m_0$ increases, the value of $\mu$
drops, much as it does in mSUGRA as we approach the focus point region. For large
$m_0$, the $\tz_1$ becomes mixed wino-higgsino, and its direct detection cross
section increases into the range which is accessible to Xenon-100. 

In Fig. \ref{fig:idd_m0_tb10}{\it b}.), we show the flux of muons from
$\nu_\mu\to \mu$ conversions at earth coming from neutralino
annihilation to SM particles within the solar core.  Here, we use the
Isajet/DarkSUSY interface for our calculations\cite{dsusy}, and require
$E_\mu >50$ GeV.  The predicted rate depends, in this case, mainly on the
sun's ability to sweep up and capture neutralinos, which depends mainly
on the {\it spin-dependent} neutralino-nucleon scattering cross section
(since in this case, the neutralinos mainly scatter from solar Hydrogen,
and there is no mass number enhancement), which is mostly sensitive to $Z^*$
exchange. The rates are again low for low $m_0$ with wino-like
neutralinos. They nearly reach the IceCube detectability level at large $m_0$ 
where the neutralinos, while remaining mainly wino-like, have picked up an increasing 
higgsino component, so that the neutralino couplings to $Z$ become large. 

In Fig. \ref{fig:idd_m0_tb10}{\it c}.), we show the expected flux of gamma
rays with $E_\gamma >1$ GeV, as required for the Fermi Gamma-ray Space
Telescope (FGST), arising from DM annihilations in the galactic core. In
this case, we see a signal rate which is flat with respect to $m_0$.
Here, the rate depends mainly on the $\tz_1\tz_1\to W^+W^-$ annihilation
cross section, which occurs via chargino exchange; since the $\tz_1$s remain mainly wino-like,
and the chargino mass hardly varies, the annihilation rate hardly varies with $m_0$.
The predictions for two halo profiles differ by over an
order of magnitude, reflecting the large uncertainty in our knowledge
of the DM density at the center of our Galaxy. Both projections are above the 
approximate reach of the FGST.

In Fig. \ref{fig:idd_m0_tb10}{\it d}.)-{\it f}.), we show the expected flux of
positrons $e^+$, antiprotons $\bar{p}$ and antideuterons $\overline{D}$ from neutralino
halo annihilations. Each of these frames show detectable rates 
by Pamela\cite{pamela} (for $e^+$s and $\bar{p}$s) and by GAPS\cite{gaps}
(for anti-deuterons). These elevated IDD rates (compared to mSUGRA\cite{bbko}
for similar $\tan\beta$ values) for anti-matter detection reflect 
the elevated rate for the $wino-wino$ annihilation into $W^+W^-$ cross section.
The halo model uncertainty for anti-matter detection is much smaller than in the $\gamma$-ray case, since
for charged particle detection, it is necessary that the anti-matter is generated
relatively close to earth, where the DM density profile is much better known.

\begin{figure}[htbp]
\begin{center}
\includegraphics[angle=0,width=0.75\textwidth]{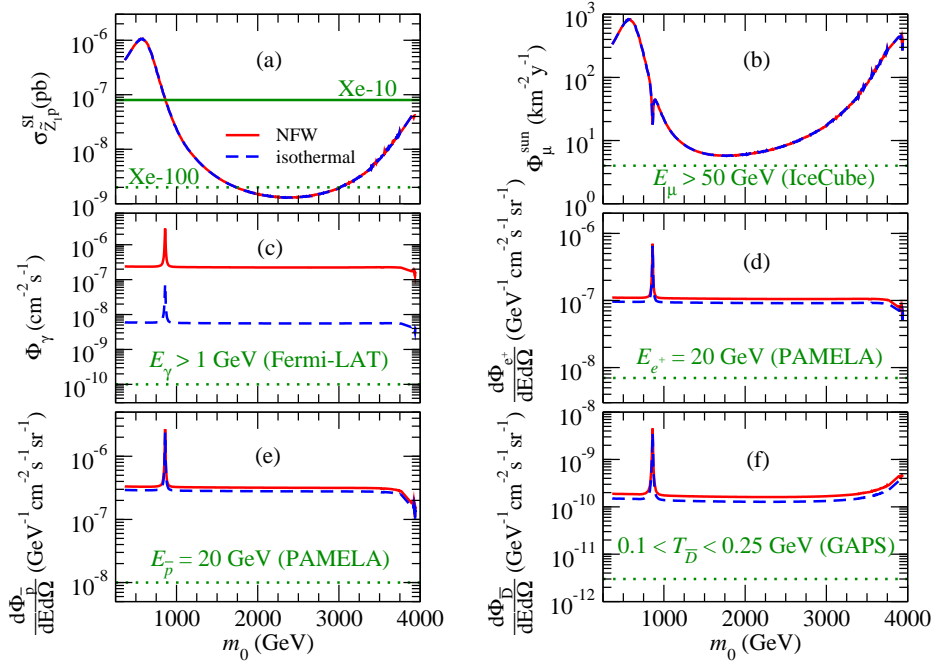}
\vspace{-.5cm}
\caption{Direct and indirect detection rates of neutralino CDM in mAMSB
vs. $m_0$, for $m_{3/2}=50$ TeV, $\tan\beta =40$ and $\mu >0$.
}
\label{fig:idd_m0_tb40}
\end{center}
\end{figure}

In Fig. \ref{fig:idd_m0_tb40}, we show rates for direct and indirect detection of wino-like WIMPs
in mAMSB versus $m_0$ with $m_{3/2}=50$ TeV, $\tan\beta =40$ and $\mu >0$.
The SI direct detection rate shown in Fig. \ref{fig:idd_m0_tb40}{\it a}.) shows a notable
enhancement at low $m_0$, and the usual enhancement at large $m_0$ due to the
increasing higgsino component of $\tz_1$. The low $m_0$ enhancement arises because the
mass of the heavy Higgs scalar $H$ has dropped with increasing $\tan\beta$, and is now quite light:
$m_H\sim 152$ GeV for $m_0=600$ GeV (compared to $m_H=1019$ GeV for the same $m_0$ with $\tan\beta =10$.
The $g\tz_1\to g\tz_1$ loop diagram via $H$ exchange is enhanced, resulting in a huge
direct detection cross section. This range of $m_0$ is already excluded by DD WIMP searches!

In Fig. \ref{fig:idd_m0_tb40}{\it b}.), we show the muon flux from mAMSB models
versus $m_0$ for $m_{3/2}=50$ TeV and $\tan\beta =40$. In this case, we again see a huge enhancement 
at low $m_0$. While normally the SD $\tz_1 p$ cross section dominates the solar accretion rate for WIMPs,
in this case, the 3 order-of-magnitude increase in SI cross section shown in frame {\it a}.) contributes
and greatly increases the solar capture rate, and hence the muon flux from the sun.
Of course, this region would already be excluded by present DD limits. We also see a curious
``anti-resonance'' effect in $\Phi_\mu$ around $m_0\sim 850$ GeV. In this case,
$m_A\sim 2m_{\tz_1}$, and neutralino annihilation is enhanced by the $A$ resonance.
Normally, for AMSB models, $\tz_1\tz_1\to VV$ ($V=W^\pm$ or $Z$) is the dominant annihilation mechanism.
But on the Higgs resonance, $\tz_1\tz_1\to b\bar{b}$ instead dominates. The energy distribution of 
neutrinos from $b$ decay is far softer than that from $W$ or $Z$ decay, leading to $\nu_\mu \to\mu$ 
conversions to lower energy muons. Since we require $E_\mu >50$ GeV for IceCube, fewer muons are detected,
and hence the anti-resonance effect.  
At large $m_0$ and $\tan\beta =40$, the muon flux is again enhanced by the increased WIMP scattering rate
via its increasing higgsino component.

In Fig. \ref{fig:idd_m0_tb40}{\it c-f}.), we see the flux of gamma rays and anti-matter versus $m_0$
at large $\tan\beta$. Here, the rate versus $m_0$ is again flat, reflecting the usually constant  
$\tz_1\tz_1$ annihilation rate into vector bosons. The exception occurs at $m_0\sim 800$ GeV, 
where annihilation through the $A$-resonance enhances the halo annihilation rate\cite{bbko}.
At large $m_0$ and $\tan\beta =40$, the $e^+$ and $\bar{p}$ detection rates drop. This is due to the
changing final state from $\tz_1\tz_1$ annihilation: at low $m_0$ it is mainly to vector bosons, leading to a hard
$e^+$ and $\bar{p}$ distribution. At large $m_0$, annihilations to $b\bar{b}$ increase and become prominent, but the
energy distribution of $e^+$ and $\bar{p}$ softens, and since we require $E_{e^+,\bar{p}}=20$ GeV, 
the detection rate drops.
In frame {\it f}.), showing the $\bar{D}$ rate, the rate actually increases at large $m_0$, 
since here we already require quite low energy $\bar{D}$s for detection, and the distribution only reflects the
increased annihilation rate.

\subsection{Indirect wino detection rates in HCAMSB}

In this subsection, we present wino-like WIMP DD and ID rates in the HCAMSB model
for $m_{3/2}=50$ TeV, versus varying $\alpha$.
As shown in Ref. \cite{hcamsb}, 
a low value of $\alpha\sim 0$ corresponds to pure anomaly-mediation, 
while large $\alpha$ gives an increasing mass $M_1$ to the hypercharge gaugino at the GUT scale. The large
value of $M_1$ pulls sparticle masses to larger values via RG evolution, with the pull 
increasing in accord with the matter state's hypercharge quantum number: 
thus-- at large $\alpha$-- we expect relatively heavy $\te_R$ states, 
but comparatively light $\tu_L$ and $\td_L$ states. 
In fact, the $U(1)$ RGE effect-- coupled with the large
$t$-quark Yukawa coupling\cite{wss}-- 
leads to relatively light, and dominantly left-, top squark states $\tst_1$ at large $\alpha$.
For very high values of $\alpha\sim 0.15-0.2$, the value of $|\mu|$ diminishes
until radiative EWSB no longer occurs (much as in the focus point region of mSUGRA).

\begin{figure}[htbp]
\begin{center}
\includegraphics[angle=0,width=0.75\textwidth]{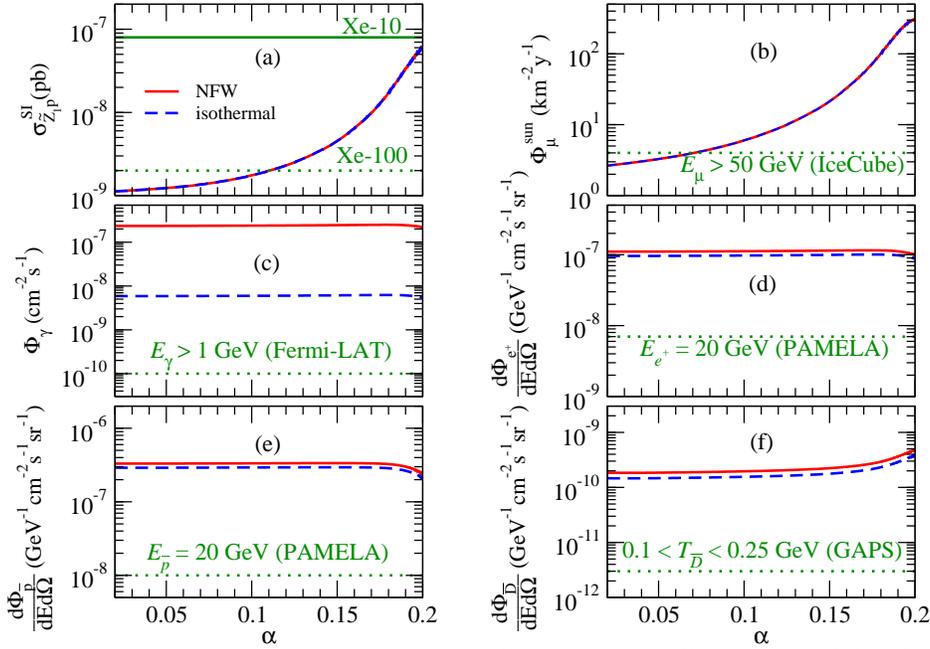}
\vspace{-.5cm}
\caption{Direct and indirect detection rates of neutralino CDM in HCAMSB
vs. $\alpha$, for $m_{3/2}=50$ TeV, $\tan\beta =10$ and $\mu >0$.
}
\label{fig:iddhc_tb10}
\end{center}
\end{figure}

In Fig. \ref{fig:iddhc_tb10}{\it a}.), we show the SI neutralino direct detection rate versus
$\alpha$ for $\tan\beta =10$. The cross section $\sigma^{SI}(\tz_1 p)$ is of order $10^{-9}$ pb at low
$\alpha$, consistent with pure wino-like neutralinos.  As $\alpha$ increases, the increasing sparticle masses
feed into $m_{H_u}^2$, diminishing the term $X_t=m_{Q_3}^2+m_{\tst_R}^2+m_{H_u}^2+A_t^2$ and leading to a 
lessened downward push by the top Yukawa coupling. Since $\mu^2\sim -m_{H_u}^2$ at the weak scale, 
the $|\mu|$ term is also diminished, leading to an increasing higgsino component of $\tz_1$. The increased
higgsino component yields an enhanced $\sigma^{SI}(\tz_1 p)$ via Higgs exchange at large $\alpha$.

In Fig.~\ref{fig:iddhc_tb10}{\it b}.), we plot the muon flux in the HCAMSB model vs. $\alpha$.
The muon flux is quite small at low $\alpha$, but at high $\alpha$, the increasing higgsino component of
$\tz_1$ leads to an increased $\sigma^{SD}(\tz_1 p)$ via $Z$-exchange. 
In Fig's~\ref{fig:iddhc_tb10}{\it c-f}.), we plot the $\gamma$-ray, $e^+$, $\bar{p}$ and $\bar{D}$ fluxes
versus $\alpha$. In these cases, the rates are large due to the large $wino-wino\to VV$ annihilation
cross section and is relatively flat with $\alpha$. At the largest $\alpha$ values, annihilation to 
$b\bar{b}$ states is enhanced, leading to diminished rates for $e^+$ and $\bar{p}$ (due to softened
energy distributions) but to a slightly increased rate for $\bar{D}$s.

\begin{figure}[htbp]
\begin{center}
\includegraphics[angle=0,width=0.75\textwidth]{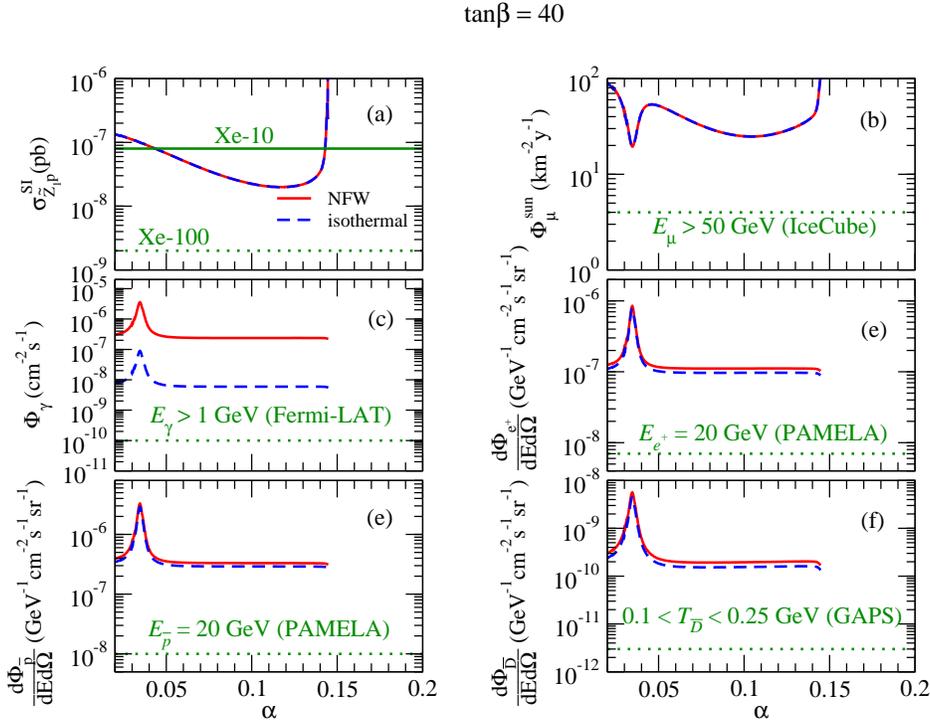}
\vspace{-.5cm}
\caption{Direct and indirect detection rates of neutralino CDM in HCAMSB
vs. $\alpha$, for $m_{3/2}=50$ TeV, $\tan\beta =40$ and $\mu >0$.
}
\label{fig:iddhc_tb40}
\end{center}
\end{figure}

In Fig.~\ref{fig:iddhc_tb40}, we show the same rates vs. $\alpha$ in the HCAMSB model, except now for
$\tan\beta =40$. The SI direct detection rate in  Fig.~\ref{fig:iddhc_tb40}{\it a}.) is enhanced 
relative to the $\tan\beta =10$ case due to the much lighter Higgs mass $m_H$ and the increased 
$b$-quark Yukawa coupling. The rate diminishes as $\alpha$ increases due to increasing squark 
and Higgs masses, 
until very high $\alpha$ is reached, and the rate is enhanced by the growing higgsino component of $\tz_1$.
The parameter space terminates above $\alpha\sim 0.15$ due to lack of REWSB.

In Fig.~\ref{fig:iddhc_tb40}{\it b}.), we see that the muon flux due to solar core annihilations
is also enhanced. In this case, the increase is again due to the large enhancement in SI
scattering cross section, which feeds into the solar accretion rate. As in Fig.~\ref{fig:idd_m0_tb40}{\it b}.),
we find an anti-resonance dip at the $\alpha$ value where $2m_{\tz_1}\sim m_A$, and WIMP
annihilations occur instead mainly into $b\bar{b}$ rather than $VV$ states.
Fig's \ref{fig:iddhc_tb40}{\it c-f}.) show the halo annihilation rates for HCAMSB at $\tan\beta =40$.
These rates are generally flat with changing $\alpha$, and do not suffer an increase compared
with low $\tan\beta$ results, since $wino-wino\to VV$ still dominates the annihilation rate.
The exception occurs at $\alpha\sim 0.035$, where $2m_{\tz_1}\sim m_A$, and halo annihilation is
enhanced by the pseudoscalar Higgs resonance.

\subsection{Indirect detection rates vs. $m_{3/2}$ for mAMSB, HCAMSB and inoAMSB}

\begin{figure}[htbp]
\begin{center}
\includegraphics[angle=0,width=0.75\textwidth]{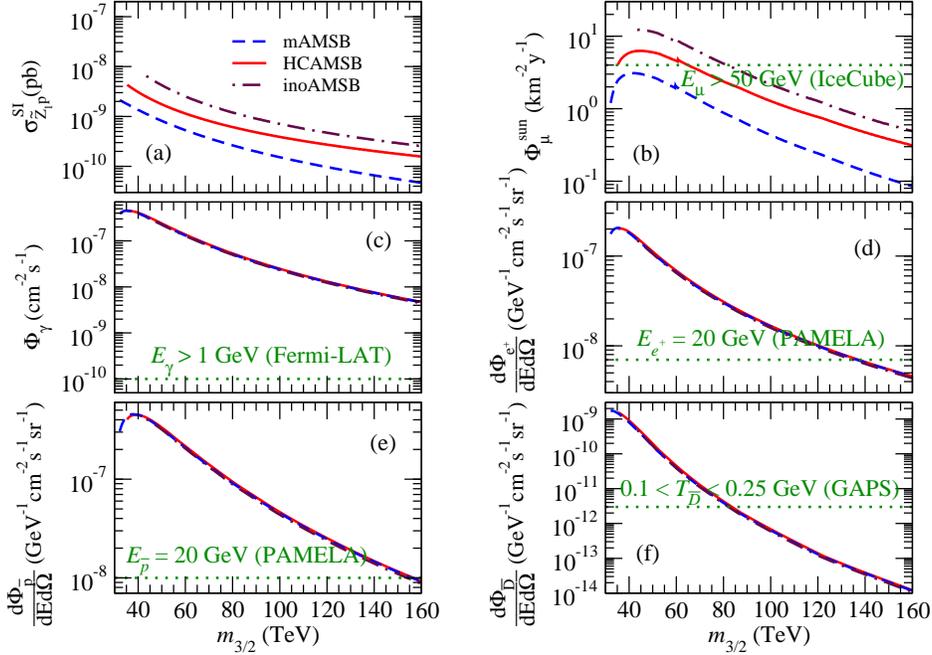}
\vspace{-.5cm}
\caption{Direct and indirect detection rates of wino CDM in mAMSB,
HCAMSB and inoAMSB models vs. $m_{3/2}$, for $\tan\beta =10$ and $\mu >0$. For mAMSB, 
we take $m_0=1$ TeV,while for HCAMSB, we take $\alpha=0.1$. 
In these plots, we adopt the NFW DM halo profile.
}
\label{fig:idd_m32_tb10}
\end{center}
\end{figure}

In Fig. \ref{fig:idd_m32_tb10}, we show direct and indirect wino DM detection rates versus
$m_{3/2}$ for $\tan\beta =10$ and $\mu >0$.  For mAMSB, we take $m_0=1$ TeV, 
while for HCAMSB, we take $\alpha=0.1$. The associated mass spectra versus $m_{3/2}$
can be found in Ref. \cite{hcamsb} for mAMSB and HCAMSB, and in Ref. \cite{inoamsb} for
the inoAMSB model. Spectra for all three models are shown in Table 1 of Ref. \cite{inoamsb}
for the case of $m_{3/2}=50$ TeV. 

In Fig. \ref{fig:idd_m32_tb10}{\it a}.), the SI direct detection cross section is shown
for all three models. In this case, we see for a given value of $m_{3/2}$, the
inoAMSB model gives the highest cross section, while mAMSB gives the lowest.  
The larger inoAMSB cross section is due in part because inoAMSB models have a smaller $\mu$
value for a given value of $m_{3/2}$, and so SI scattering via Higgs exchange (which
involves a product of higgsino and gaugino components) is enhanced.

In Fig. \ref{fig:idd_m32_tb10}{\it b}.), we show the relative rates for indirect wino detection 
due to WIMP annihilation into $\nu_\mu$ states in the solar core, 
with subsequent muon detection from $\nu_\mu\to\mu$ 
conversions in Antarctic ice, as might be seen by IceCube. We require $E_\mu >50$ GeV.
The muon flux is mainly related to the spin-dependent direct detection rate, which
enters the sun's ability to capture WIMPs.
Here again, inoAMSB yields the highest rates, and mAMSB the lowest. This follows the
relative values of $\mu$ in the three models: low $\mu$ in inoAMSB leads to a larger
higgsino component of $\tz_1$, and an increased SD scattering rate via $Z^*$ exchange.
A rough reach of the IceCube detector is shown, and indicates that the low $m_{3/2}$
portion of parameter space of inoAMSB and HCAMSB may be accessible to $\nu_\mu\to\mu$  searches.

In Fig. \ref{fig:idd_m32_tb10}{\it c-f}.), we show the ID rates for detection of $\gamma$, $e^+$s, 
$\bar{p}$s and $\bar{D}$s, for the energy ranges indicated on the plots. 
All these plots adopt the NFW halo profile.
In all these cases, 
all three models yield almost exactly the same detection rates for a given value of 
$m_{3/2}$. This is due to the dominance of $\tz_1\tz_1\to VV$ halo annihilations, 
which mainly depend on the gaugino component of $\tz_1$, which is nearly all wino-like.
The rough reach of Fermi-LAT, Pamela and GAPS is shown for reference. The high rates for
wino halo annihilations should yield observable signals. As mentioned previously, 
Kane {\it et al.} promote wino-like WIMPs as a source of the Pamela anomaly\cite{kane}.
In this case, a large $\bar{p}$ signal should be seen as well, although the Pamela
$\bar{p}$ rate seems to agree with SM background projections.

\begin{figure}[htbp]
\begin{center}
\includegraphics[angle=0,width=0.75\textwidth]{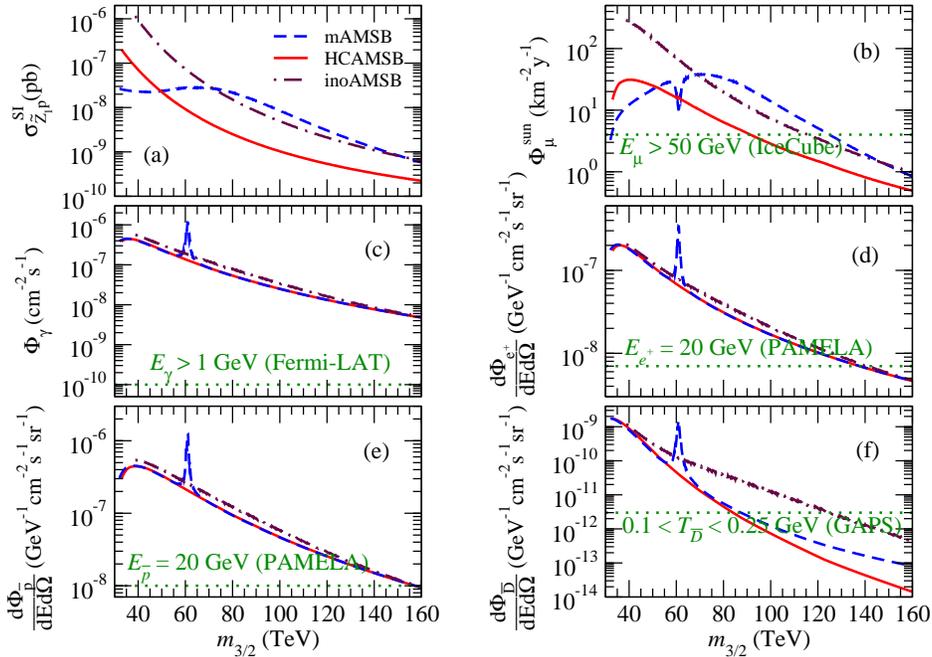}
\vspace{-.5cm}
\caption{Direct and indirect detection rates of wino CDM in mAMSB,
HCAMSB and inoAMSB 
vs. $m_{3/2}$, for $\tan\beta =40$ and $\mu >0$. For mAMSB, 
we take $m_0=1$ TeV,while for HCAMSB, we take $\alpha=0.1$.
In these plots, we adopt the NFW DM halo profile.
}
\label{fig:idd_m32_tb40}
\end{center}
\end{figure}

In Fig. \ref{fig:idd_m32_tb40}, we show direct and indirect wino detection rates
versus $m_{3/2}$ as in Fig. \ref{fig:idd_m32_tb10}, 
except now for a large value of $\tan\beta =40$.
In frame {\it a}.), we see that the SI direct detection rates are all elevated with respect 
to the $\tan\beta =10$ case. The well-known large $\tan\beta$ enhancement\cite{dn,bb}
occurs due to enhanced Higgs exchange contributions, where now the 
value of $m_H$ is lower and the $b$-quark Yukawa coupling is larger.
For low $m_{3/2}$, the inoAMSB model has the smallest value of $m_H$ and
the lowest value of $\mu$ for a given $m_{3/2}$ value, and thus the highest 
value of $\sigma^{SI}(\tz_1 p)$. As $m_{3/2}$ increases, the value of $m_H$
increases for inoAMSB and HCAMSB, while it actually {\it decreases} for
mAMSB. Thus, for $m_{3/2}\agt 75$ TeV, the mAMSB model yields the highest
value of $\sigma^{SI}(\tz_1 p)$.

In Fig. \ref{fig:idd_m32_tb40}{\it b}.), we show the muon flux for IceCube due to wino annihilation
in the solar core. Here again, the rates are elevated compared to the $\tan\beta =10$ case.
The inoAMSB model yields the highest flux at low $m_{3/2}$, 
since it has the lowest $\mu$ value, and the highest higgsino component, which enters
into the $Z^*$ exchange diagram for $q\tz_1$ scattering. 
At $m_{3/2}\sim 60$ TeV in the mAMSB model, we obtain
$2m_{\tz_1}\sim m_A$, and the solar core annihilations mainly proceed to $b\bar{b}$
states instead of $VV$, which diminishes the muon energy distribution and hence the 
detection rate for $\mu$s with $E_\mu >50$ GeV. At higher $m_{3/2}$ values, 
the resonance is passed, and annihilation once again proceeds dominantly into $VV$.
For high $m_{3/2}$, the mAMSB model yields the highest muon flux, due to its elevated
value of $\sigma^{SI}(\tz_1 p)$.

In Fig. \ref{fig:idd_m32_tb40}{\it c-f}.), we find relatively little change in halo annihilation rates due
to an increase in $\tan\beta$, since the annihilations mainly proceed via $wino-wino\to VV$, which 
depends mainly on gauge couplings. 
The exception occurs in the mAMSB model, where we do get
the resonance enhancement of halo annihilations when $2m_{\tz_1}\sim m_A$. We also obtain
some $\tan\beta$ enhancement of the $\bar{D}$ detection rate for inoAMSB and mAMSB at large
$m_{3/2}$ because in these cases the $\tz_1\tz_1\to b\bar{b}$ annihilation rate, which
does receive $\tan\beta$ enhancement, contributes to the detection of rather low energy
$\bar{D}$s.

\section{Discussion and conclusions}
\label{sec:conclude}

In this paper, we have investigated aspects of cold dark matter in three models
of anomaly mediation: mAMSB, HCAMSB and inoAMSB. Typically, each gives rise to a
wino-like lightest neutralino, unless very high values of $m_0$ (for mAMSB) or
$\alpha$ (for HCAMSB) are used, in which case the $\tz_1$ becomes a mixed
wino-higgsino state.
In this class of models with a wino-like $\tz_1$, the thermal abundance of
neutralino CDM is well below measured values, unless $m_{\tz_1}\agt 1300$ GeV.
We discuss four ways to reconcile the predicted abundance of CDM with experiment:
\begin{enumerate}
\item enhanced neutralino production via scalar field ({\it e.g.} moduli) decay, 
\item enhanced neutralino production via gravitino decay, where gravitinos may arise
thermally, or by moduli or inflaton decay, 
\item enhanced neutralino production via heavy
axino decay, and 
\item neutralino decay to axinos, where the bulk of CDM comes from a 
mixture of vacuum mis-alignment produced axions and thermally produced axinos.
\end{enumerate}
Cases 1 and 2 should lead to a situation where all of CDM is comprised of wino-like WIMPs; 
they will be very hard, perhaps impossible, to tell apart.
Case 3 would contain a mixture of axion and wino-like WIMP CDM. It is a scenario where
it is possible that both a WIMP and an axion could be detected. Case 4 predicts no
direct or indirect detection of WIMPs, but a possible detection of relic axions.
It is important to note that more than one of these mechanisms may occur at once: for instance, 
we may gain additional neutralino production in the early universe from moduli, gravitino and axino
decay all together.

In Sec. \ref{sec:winodet}, we presented rates for direct and indirect detection of relic wino-like
WIMPs. The SI direct detection cross sections are bounded from below. Ultimately, ton-scale
noble liquid or SuperCDMS experiments should probe out to $m_{\tz_1}\sim 500$ GeV, which would
exceed the 100 fb$^{-1}$ reach of LHC; a non-observation of signal would put enormous stress
on AMSB-like models as new physics. We also evaluated SD direct detection: current experiments
have little reach for AMSB-like models, although IceCube DeepCore and possibly COUPP upgrades may probe
more deeply.

We also presented indirect WIMP detection rates for all three AMSB models. The IceCube experiment
has some reach for WIMPs from AMSB models, especially at high $\tan\beta$ or when the
$\tz_1$ picks up a higgsino component. We noted an interesting inverse resonance effect in the
muon flux detection rate, caused by transition from solar core annihilations to $VV$ states, 
to annihilations to mainly $b\bar{b}$ states. 
The detection of $\gamma$s, $e^+$s, $\bar{p}$s and $\bar{D}$s are all elevated in AMSB-like models
compared to mSUGRA, due to the high rate for $\tz_1\tz_1\to VV$ annihilation in the galactic
halo. The results do depend on the assumed halo profile, especially for $\gamma$-ray detection
in the direction of the galactic core. Generally, if a signal is seen in the $e^+$ channel, 
then one ought to be seen in the $\bar{p}$ channel, and ultimately in the $\gamma$, 
$\bar{D}$ (if/when GAPS flies) or direct detection channel. In addition, a sparticle 
production signal should ultimately be seen at LHC, at least for $m_{\tg}\alt 2400$ GeV, once
100 fb$^{-1}$ of integrated luminosity is accrued.

As a final remark, we note here that the dark matter detection signals all provide
complementary information to that which will be provided by the CERN LHC.
At LHC, each model-- mAMSB, HCAMSB and inoAMSB-- will provide a rich assortment of
gluino and squark cascade decay signals which will include multi-jet plus multi-lepton
plus missing $E_T$ events. In all cases, 
the wino-like lightest neutralino state will be signaled
by the well-known presence of highly ionizing tracks (HITs) from 
quasi-stable charginos with track length of order
{\it cm}s, before they decay to soft pions plus a $\tz_1$. 
It is noted in Ref's \cite{hcamsb} and \cite{inoamsb} that the three models 
should be distinguishable at LHC by the differing opposite-sign/same flavor
dilepton invariant mass distributions. In the case of mAMSB, with 
$m_{\tell_L}\simeq m_{\tell_R}$, we expect a single mass edge from
$\tz_2\to\ell\tell_{L,R}\to\ell^+\ell^-\tz_1$ decay. In HCAMSB, the sleptons
are rather heavy, and instead $\tz_2\to\tz_1 Z$ occurs at a large rate, 
leading to a bump in $m(\ell^+\ell^-)\sim M_Z$, upon a continuum
distribution. In inoAMSB, with 
$m_{\tz_2}>m_{\tell_{L,R}}$, but with $\tell_L$ and $\tell_R$ split in mass
(due to different $U(1)_Y$ quantum numbers), a characteristic {\it double mass edge}
is expected in the $m(\ell^+\ell^-)$ invariant mass distribution.

\section*{Acknowledgments}
We thank Shanta de Alwis for comments on the manuscript.
This work was supported in part by the U.S.~Department of Energy.
%

%
\end{document}